\documentclass{article}
\usepackage{amsmath}
\usepackage{amsfonts}
\usepackage{amssymb}
\usepackage{authblk}
\usepackage{color}
\usepackage{graphicx} 
\usepackage{soul}
\usepackage{physics}
\usepackage{geometry}
\usepackage{hyperref}
\usepackage{mathtools}
\usepackage{titlesec}

\titleclass{\subsubsubsection}{straight}[\subsubsection]
\newcounter{subsubsubsection}[subsubsection]
\renewcommand\thesubsubsubsection{\thesubsubsection.\arabic{subsubsubsection}}

\titleformat{\subsubsubsection}
  {\normalfont\normalsize\itshape}{\thesubsubsubsection}{1em}{}
\titlespacing*{\subsubsubsection}{0pt}{1.0ex}{0.5ex}

\title{
Characterization of Phase Transitions in a Lipkin–Meshkov–Glick Quantum Brain Model}
\author[1,*]{Elvira Romera}
\author[2]{Joaqu\'in  J. Torres}
\affil[1]{Departamento de F\'{\i}sica At\'omica, Molecular y Nuclear and 
	Instituto Carlos I de F\'\i sica Te\'orica y Computacional, Universidad de Granada, Fuentenueva s/n, 18071 Granada,
	Spain}
\affil[2]{Departamento de Electromagnetismo y Física de la Materia and 
	Instituto Carlos I de F\'\i sica Te\'orica y Computacional, Universidad de Granada, Fuentenueva s/n, 18071 Granada,
	Spain}
\affil[*]{Corresponding Author: eromera@ugr.es}

\begin{document}
\maketitle

\begin{abstract}
In this work we analyze the emergence of phase transitions in a quantum brain model inspired by the Lipkin--Meshkov--Glick framework, where biologically motivated synaptic feedback modulates the collective interaction in a nonlinear and state-dependent manner. We demonstrate that incorporating this retroactive mechanism substantially reshapes the phase structure, yielding an expansion of the paramagnetic phase at the expense of the ferromagnetic phases relative to the feedback-free scenario. This effect is markedly enhanced in the presence of a longitudinal field, as the feedback couples directly to the longitudinal magnetization, leading to an appreciable displacement of the critical boundaries. We characterize the ensuing transitions from a phase-space perspective by means of the ground-state Husimi distribution and the Wehrl entropy, which provide a robust diagnosis of qualitative changes in localization and enable a quantitative assessment of feedback-induced deformations of the phase diagram.  Additionally, we perform an explicit dynamical analysis based on mean-field equations for the collective-spin orientation self-consistently coupled to the synaptic dynamics, which reproduces with high fidelity the quantum time evolution of collective observables for the protocols considered. Overall, these findings substantiate the suitability of this quantum brain model as a controlled theoretical framework for elucidating how synaptic plasticity mechanisms can parametrically tune and reshape collective criticality.
\end{abstract}

\section{Introduction}
State-Dependent Feedback in a Quantum Brain Model
The Lipkin--Meshkov--Glick (LMG) model has recently been introduced as a natural substrate for biologically inspired quantum neural networks, in which neuronal populations are encoded as fully connected qubits governed by LMG-type collective dynamics and stabilized through synaptic-efficacy feedback implementing activity-dependent homeostatic control \cite{TorresRomera2026}. Owing to its capacity to encode quantum many-body collectivity in a compact yet remarkably expressive form, the LMG Hamiltonian provides a natural foundation for the present development. First introduced in nuclear physics \cite{Lipkin1965a,Lipkin1965b,Lipkin1965c,RingSchuck1980} as a controlled arena to probe particle--particle and particle--hole correlations in neutron--proton systems and to benchmark approximate many-body approaches, the model has since attained a much broader relevance. In condensed-matter settings it furnishes effective descriptions of Bose--Einstein condensates and Josephson-junction dynamics \cite{MilburnEtAl1997,Leggett2001,MicheliEtAl2003}, whereas in quantum optics it underpins both the metrological assessment of spin-squeezed states and the engineered generation of multipartite entanglement, in the presence of external fields as well as for field-free quadratic collective-spin Hamiltonians \cite{KitagawaUeda1993,JurcevicEtAl2014,RichermeEtAl2014,PhysRevA.108.023722}. Consequently, the LMG model has become a canonical paradigm for collectively interacting two-level systems: it supports first-, second-, and third-order quantum phase transitions \cite{Castanos2006PRB,DusuelVidal2004}, and its phase structure has been rigorously elucidated through phase-space delocalization measures complemented by entanglement-entropy diagnostics \cite{NagyRomera2012,RomeraCalixtoNagy2012,CalixtoRomeraDelReal2012,CastanosCalixtoPerezBernalRomera2015,CalixtoCastanosRomera2017_JSTAT}.

In classical and biologically inspired neural networks models, synapses occupy a central role as nonlinear elements governing information transfer \cite{Amit2012ModelingBrainFunction}: transmission relies not only on pairwise processes such as neurotransmitter release and recycling, but also on higher-order mechanisms, prominently including astrocyte-mediated modulation of synaptic efficacy \cite{Menesse2025AstrocyteControl}. Experimental evidence further establishes that synaptic efficacy is intrinsically dynamic and activity dependent, decreasing through synaptic depression or increasing via facilitation under presynaptic stimulation \cite{Tsodyks1998}; such modulation entails substantial computational consequences \cite{Torres2013}, shaping memory capacity \cite{Torres2002,Mejias2009}, dynamic memory formation \cite{Pantic2002,Torres2007}, and stochastic resonance in weak-signal processing \cite{Mejias2011}, and it may additionally perturb excitation--inhibition balance, thereby precipitating abrupt surges of excitatory activity and emergent brain rhythms with distinct informational content \cite{Pretel2021,Menesse2024}. Yet a considerable fraction of the recent quantum-neural-network literature idealizes the problem by replacing binary neurons with qubits endowed with simplified couplings, despite the functional salience of synaptic dynamics; in parallel, theoretical frameworks have been developed for both single quantum neurons \cite{Cao2017QuantumNeuron,Tacchino2019ArtificialNeuron,Kristensen2021SpikingQuantumNeuron} and network-level architectures, including quantum perceptrons \cite{Pechal2021QuantumPerceptron,Wiebe2016QuantumPerceptron} and Hopfield networks \cite{Rotondo2018}, whose computational properties and storage capacity have been quantified in recent work \cite{Torres2024}, with the overarching aim of determining whether quantum analogues can outperform classical networks in pattern recognition, classification, and memory storage while retaining biologically inspired principles \cite{Torres2022}. Along these lines, a quantum-synapse framework based on activity-dependent qubit couplings has been advanced to examine how synaptic depression reshapes qubit interactions and entanglement; however, its formulation---restricted to a two-qubit setting and relying on a detailed microscopic treatment---renders scaling to large networks impractical due to the rapid growth of the relevant state-space dimensionality with system size \cite{Torres2022}.

On the other hand, it has been established that phase-space (quasi-)probability representations provide a systematic route to characterize quantum states and their collective behavior, with the Wigner and Husimi functions playing a particularly prominent role. The Husimi distribution is especially advantageous because, unlike the Wigner function, it is strictly non-negative, which facilitates both visual inspection and quantitative diagnostics in contexts ranging from metal--insulator transitions \cite{njp} to quantum chaos \cite{leboeuf}. Within this broader program, a sequence of recent studies has consolidated the Husimi function as a sensitive probe of quantum phase transitions (QPTs) in paradigmatic many-body models most notably the Dicke, Vibron, and LMG models, through uncertainty-based indicators such as the inverse participation ratio and R\'enyi--Wehrl entropies \cite{husimi1,husimi2,husimi3,Romera2014,Castanos2015}. Concretely, for the Dicke model \cite{husimi1} numerical analyses were combined with analytical variational approximations. It was found that the inverse participation ratio and the Wehrl entropy provide clear signatures of the transition. Within the variational framework, this description is complemented by a characterization in terms of the zeros of the Husimi distribution. Similarly, in the two-dimensional $U(3)$ vibron model for molecules of size $N$ \cite{husimi2}, these same phase-space indicators unambiguously identify the (shape) transition between linear and bent configurations. The numerical results are supported by a parity-adapted $U(3)$  coherent-state variational approach. This approach attains the minimum Wehrl entropy in the strictly linear phase, in agreement with a generalized Wehrl--Lieb conjecture. Again, the zeros of the distribution provide an additional characterization.
 Along closely related lines, it has also been shown that entropic measures constructed from marginals of the Husimi function can efficiently identify and characterize QPTs in the Dicke and LMG models \cite{husimi1,Romera2026HusimiLMG}.

In this work, we analyze both the static ground-state properties and a quantum--classical dynamical extension of the LMG model, where the interaction strength is modulated by a classical feedback variable and which recently has been proposed as a suitable framework for a Quantum Brain model easily implemented in the actual developing quantum platform as the current quantum computers \cite{TorresRomera2026}. Our aim is to study in deep the presence of phase transitions on such quantum brain and explore the computational properties of the emerging phases. We use the Wehrl entropy $W$ as a quantitative measure of phase-space delocalization. We also dynamically analyze the behavior of the system through the whole phase diagram and in particular around the emerging phase transitions.To do so, we directly compare the exact quantum evolution of the system with an original dynamical mean-field description based on coherent states that characterize the collective dynamics of the quantum brain. This comparison allows us to assess the validity of the approximation and to determine the role of quantum correlations beyond the semiclassical regime.


\section{Models and Methods}
\subsection{Anisotropic LMG Hamiltonian}

We consider the fully connected $XY$ model in the presence of an external field along the $z$ direction,
\begin{equation}
H_{XY}= -\frac{\lambda}{N}\sum_{i<j}\Big(\sigma_i^{x}\sigma_j^{x}+\gamma\,\sigma_i^{y}\sigma_j^{y}\Big)
-\frac{h}{2}\sum_{i=1}^{N}\sigma_i^{z}.
\end{equation}
This Hamiltonian encodes an exchange-type coupling among spin degrees of freedom in the $x\!-\!y$ plane, with an anisotropy controlled by $\gamma$. The interaction is inherently collective and admits a mean-field-like reformulation in terms of total-spin operators. To this end, we introduce the collective spins
\begin{equation}
J_{\alpha}=\frac{1}{2}\sum_{i=1}^{N}\sigma_i^{\alpha},\qquad \alpha=x,y,z,
\end{equation}
and employ the identity
\begin{equation}
\sum_{i<j}\sigma_i^{\alpha}\sigma_j^{\alpha}=2J_{\alpha}^{2}-\frac{N}{2}.
\end{equation}
Up to an additive constant that merely shifts the zero of energy, this yields an equivalent Hamiltonian expressed solely in terms of collective operators. Absorbing the resulting numerical prefactors into the coupling $\lambda$, one arrives at the anisotropic Lipkin--Meshkov--Glick (LMG) Hamiltonian,
\begin{equation}
H_{LMG}= -\frac{\lambda}{N}\left(J_{x}^{2}+\gamma J_{y}^{2}\right)-hJ_{z}.
\end{equation}
Here, $N$ denotes the total number of spins (or qubits), and the $1/N$ prefactor ensures an extensive energy scaling; $\lambda$ sets the overall strength of the collective (effectively two-body) interaction; $\gamma$ quantifies the anisotropy between the $x$ and $y$ channels by weighting $J_y^{2}$ relative to $J_x^{2}$; and $h$ is an effective external field coupling linearly to $J_z$, thereby favouring polarization along $z$ in competition with the transverse $x\!-\!y$ interaction.

The total angular momentum $\vec{J}^{\,2}=j(j+1)$ and the total number of particles $N=2j$ are conserved; moreover, we take $\hbar=1$. The Hamiltonian commutes with the parity operator
\[
\hat{P}=e^{i\pi(J_z+j)},
\]
for a fixed $j$.

We shall focus on non-trivial regimes of $H$ that give rise to quantum phase transitions (QPTs) of first, second, and third order, according to the standard Ehrenfest classification. A comprehensive classification of the critical points of the associated energy surface has been carried out within the framework of catastrophe theory; see Ref.~\cite{Castanos2005a}  .

\subsubsection{Semiclassical description via coherent states}
\label{sec:semi}
A natural semiclassical description in the thermodynamic limit is obtained from spin-$j$ coherent states (Refs.~\cite{PerelomovBook,GilmoreBook}). They are defined as
\begin{equation}
\ket{\zeta} \;=\; (1+\abs{\zeta}^2)^{-j}\sum_{m=-j}^{j}\binom{2j}{j+m}^{1/2}\,\zeta^{\,j+m}\,\ket{j,m},
\label{eq:coherent}
\end{equation}
with the stereographic parametrization
\[
\zeta=\tan\left(\frac{\theta}{2}\right)e^{i\phi},
\]
where $\binom{2j}{j+m}$ denotes a binomial coefficient.

These states are commonly regarded as the most ``classical'' states of a collective spin: they are localized on the Bloch sphere, they saturate the appropriate $SU(2)$ uncertainty relations (minimum-uncertainty states in the group-theoretic sense), and their expectation values reproduce a classical spin vector of length $j$ oriented by the angles $(\theta,\phi)$~\cite{Arecchi1972,GilmoreBook}. Since the LMG Hamiltonian is written solely in terms of the collective operators $J_x$, $J_y$, and $J_z$, the dynamics can be consistently restricted to the fully symmetric sector, which for $N$ spin-$1/2$ constituents corresponds to a single collective spin of length $j=N/2$~\cite{BengtssonZyczkowski2017}.

The set $\{\ket{\zeta}\}$ is overcomplete and satisfies the resolution of the identity
\begin{equation}
\mathbf{1}=\int_{\mathbb{R}^{2}}\ket{\zeta}\bra{\zeta}\,d\mu(\zeta).
\tag{3}
\label{eq:resolution}
\end{equation}
The corresponding invariant measure reads
\begin{equation}
d\mu(\zeta)=\frac{2j+1}{\pi}\frac{d^{2}\zeta}{(1+\abs{\zeta}^{2})^{2}}
=\frac{2j+1}{4\pi}\sin\theta\,d\theta\,d\phi,
\tag{4}
\label{eq:measure}
\end{equation}
where $d^{2}\zeta \equiv d\,\mathrm{Re}(\zeta)\, d\,\mathrm{Im}(\zeta)$.

Evaluating the LMG Hamiltonian on spin coherent states yields the associated energy surface $\mathcal{E}(\theta,\phi)$ (which will be addressed in the following). The qualitative phase structure and the onset of quantum phase transitions can then be inferred from the organization of its stationary points (minima, maxima, and saddles) ~\cite{Castanos2006PRB}.

The energy surface is obtained by evaluating the Hamiltonian on coherent states. We use calligraphic symbols for coherent-state expectation values, so one has
\[
\mathcal{J}_\alpha \equiv \mel{\zeta}{J_\alpha}{\zeta},
\qquad \alpha=x,y,z.
\]
In particular, this results in
\begin{equation}
\mathcal{J}_x=j\,\sin\theta\,\cos\phi,
\qquad
\mathcal{J}_y=j\,\sin\theta\,\sin\phi,
\qquad
\mathcal{J}_z=j\,\cos\theta.
\label{eq:Jsymbols}
\end{equation}
For a better characterization of the emerging phases it is convenient also to consider the magnetization variables defined as $m_\alpha=\mathcal{J}_\alpha/j.$

In the semiclassical (large-$j$) limit one may approximate $\mel{\zeta}{J_\alpha^2}{\zeta}\simeq \mathcal{J}_\alpha^2$. The mean energy per particle is then defined as
\[
\mathcal{E}(\theta,\phi)=
\frac{\mel{\zeta}{H}{\zeta}}{2j\,\braket{\zeta}{\zeta}}.
\]
Using $N=2j$, we obtain the explicit expression
\begin{equation}
\mathcal{E}(\theta,\phi)=
-\frac{\lambda}{4}\sin^2\theta\Big(\cos^2\phi+\gamma\sin^2\phi\Big)
-\frac{h}{2}\cos\theta.
\label{eq:Esurface}
\end{equation}
It is convenient to introduce
\[
f(\phi)=\cos^2\phi+\gamma\sin^2\phi
=1+(\gamma-1)\sin^2\phi,
\]
so that
\[
\mathcal{E}(\theta,\phi)= -\frac{\lambda}{4}\sin^2\theta\,f(\phi)
-\frac{h}{2}\cos\theta.
\]

The critical points of the energy surface $E(\theta,\phi)$ are determined by imposing the stationary conditions
\begin{equation}
\left\{
\begin{aligned}
\frac{\partial \mathcal{E}}{\partial \theta} &= 0,\\[4pt]
\frac{\partial \mathcal{E}}{\partial \phi} &= 0,
\end{aligned}
\right.
\qquad\Longrightarrow\qquad
\begin{cases}
\theta = \theta_c(\lambda,\gamma,h),\\
\phi = \phi_c(\lambda,\gamma,h),
\end{cases}
\tag{8}
\label{eq:criticalsystem}
\end{equation}
i.e., the critical angles $(\theta_c,\phi_c)$ are defined as functions of the control parameters. Within this semiclassical picture, these stationary points represent the relevant phase-space configurations (minima, maxima, or saddle points) and therefore encode the qualitative structure of the energy landscape. We will use this procedure to obtain the emerging phases in our quantum brain model in a straightforward maner (see Results Section below).

\subsubsection{Husimi function and R\'enyi--Wehrl entropies}
In order to carry out a phase-space analysis of quantum phase transitions in the LMG model (and, analogously, in the quantum brain model), we employ the Husimi function $Q_{\psi}$ associated with the ground state $\ket{\psi}$. It is defined as the squared modulus of the overlap between $\ket{\psi}$ and an arbitrary spin-coherent state $\ket{\zeta}$,
\begin{equation}
Q_{\psi}(\zeta)=\left|\braket{\zeta}{\psi}\right|^{2},
\tag{19}
\label{eq:HusimiDef}
\end{equation}
Expanding $\ket{\psi}$ in the Dicke basis $\{\ket{j,m}\}$ as $\ket{\psi}=\sum_{m=-j}^{j}c_m\,\ket{j,m}$, one obtains
\begin{equation}
Q_{\psi}(\zeta)
=\sum_{m,m'=-j}^{j}c_{m}\,\bar{c}_{m'}\,\varphi^{\,j}_{m}(\zeta)\,
\overline{\varphi^{\,j}_{m'}(\zeta)}.
\tag{20}
\label{eq:HusimiExp}
\end{equation}
Here $c_m$ denote the expansion coefficients of $\ket{\psi}$ in the $\ket{j,m}$ basis and
\begin{equation}
\varphi^{\,j}_{m}(\zeta)=\braket{j,m}{\zeta}
=
\frac{\sqrt{\binom{2j}{j+m}}\;\zeta^{\,j+m}}{(1+\abs{\zeta}^{2})^{j}}
\tag{21}
\label{eq:phi}
\end{equation}
is the coherent-state representation of the Dicke states. By construction, $Q_{\psi}(\zeta)\ge 0$
and it is normalized with respect to the invariant measure $d\mu(\zeta)$,
\begin{equation}
\int_{\mathbb{R}^{2}} Q_{\psi}(\zeta)\,d\mu(\zeta)=1,
\tag{22}
\label{eq:HusimiNorm}
\end{equation}
where $d\mu(\zeta)$ is given in Eq.~\eqref{eq:measure}.

In the following, we introduce the R\'enyi--Wehrl entropies \cite{Wehrl1978}, which have been shown to provide a very useful tool for analyzing quantum phase transitions in a variety of systems~\cite{Castanos2006PRB,njp,husimi1,husimi2,husimi3}.
.

\begin{equation}
W_{\nu}=-\frac{1}{1-\nu}\ln M_{\nu},
\tag{32}
\label{eq:RWentropy}
\end{equation}
where
\begin{equation}
M_{\nu}=\int_{\mathbb{R}^{2}}\left(Q_{\psi}(\zeta)\right)^{\nu}\,d\mu(\zeta)
\tag{33}
\label{eq:HusimiMoment}
\end{equation}
is the $\nu$-th moment of the Husimi function. In particular, in the limit $\nu\to 1$ one recovers the Wehrl entropy,
\begin{equation}
W=-\int_{\mathbb{R}^{2}} Q_{\psi}(\zeta)\,\ln\!\big(Q_{\psi}(\zeta)\big)\,d\mu(\zeta).
\tag{34}
\label{eq:WehrlEntropy}
\end{equation}


Two limiting phase-space configurations are particularly informative. Since the invariant measure satisfies $\int_{\mathbb{R}^2} d\mu(\zeta)=2j+1$, the uniform Husimi distribution $Q(\zeta)=1/(2j+1)$ is properly normalized and yields maximal entropy values, namely $W=\ln(2j+1)$ and $W_{\nu}=\ln(2j+1)$ for $\nu>1$. At the opposite extreme, $SU(2)$ coherent states are the most localized phase-space packets: they minimize the Wehrl entropy (\cite{Wehrl1978}) and attain the sharp lower bound $W\geq W_{\min}=2j/(2j+1)$ proved for spin coherent states in Ref.~\cite{LiebSolovej2014}.

The lower bounds of these entropies are ultimately rooted in the fact that $SU(2)$ coherent states are, in a precise sense, the \emph{most localized} states in phase space (they minimize the Wehrl entropy), whereas the upper bounds are attained by Husimi distributions that are (approximately) uniform over the Bloch sphere, corresponding to maximal delocalization. In particular, for $\nu>1$ the pointwise constraint $0\leq Q_{\psi}(\zeta)\leq 1$ immediately implies the general bound $W_{\nu}\geq 0$. Combining these observations with the sharp Wehrl-minimization result for spin coherent states yields finite-$j$ bounds of the form $\frac{2j}{2j+1}\leq W\leq \ln(2j+1)$ and $0\leq W_{\nu}\leq \ln(2j+1)$ (for $\nu>1$), which in the thermodynamic limit translate into $W\gtrsim 1$ and $W_{\nu}\lesssim \ln(2j)$ up to subleading $\mathcal{O}(1/j)$ corrections. This geometric interpretation of $W$ and $W_{\nu}$ as quantitative measures of phase-space localization/delocalization is standard.~\cite{Wehrl1978,LiebSolovej2014,GnutzmannZyczkowski2001}

On the other hand, if the state is a normalized coherent superposition of $p$ well-separated packets on the Bloch sphere  and Husimi interference fringes are negligible, the density $Q_{\psi_p}$ is well approximated by a sum of $p$ essentially disjoint lobes of equal weight. In this regime the entropy exhibits a characteristic additive shift, $W_{\nu}[\psi_{p}]\simeq W_{\nu}[\zeta_{0}] + \ln p$ (and in particular $W[\psi_{p}]\to 1+\ln p$ as $j\to\infty$), reflecting that the state effectively occupies $p$ coherent ``cells'' in phase space. This additivity under well-resolved coherent components underlies the usefulness of $W_{\nu}$ as a sensitive diagnostic of state fragmentation, for instance in the vicinity of quantum phase transitions.~\cite{GnutzmannZyczkowski2001,Romera2014,Castanos2015}

\subsection{Quantum brain model}
Recently, we have introduced  a quantum brain model based on the LMG model \cite{TorresRomera2026}. One of the main features of such system is the presence of biologically motivated synaptic feedback affecting the collective interaction among qubtis. The dynamic of  such system is given by equations
\begin{align}
&i \frac{d}{dt} |\psi(t)\rangle = H(r(t)) |\psi(t)\rangle, \nonumber\\
&\frac{dr(t)}{dt} = \frac{1-r(t)}{\tau_r} - U(t) r(t) E(t),\label{brainmodel}\\
&\frac{dU(t)}{dt} = \frac{\mathcal{U}-U(t)}{\tau_f} + \mathcal{U}[1-U(t)]  E(t) ,\nonumber
\end{align}
with
\begin{equation}
H(r(t)) = -\frac{\lambda_0 r(t)}{N} \left( J_x^2 + \gamma J_y^2 \right) - h J_z
\label{brainham}
\end{equation}
and $E(t)=(1+m_z(t))/2.$ Here, $\mathcal{U},$ $\tau_r$ and $\tau_f$ are parameters controlling the dynamics of the synaptic feedback and have some biological origin related with the trafficking, release and recycling of neurotransmitters vesicles in actual brains \cite{Tsodyks1998}. The second and third equations of (\ref{brainmodel}) are related to the  short-term plasticity mechanisms of synaptic depression and facilitation respectively. More precisely the second equation introduce synaptic fatigue when the presynaptic activity is high acting as a filter and the third equation transiently enhances the  synaptic response by increasing the parameter $U(t).$  This synaptic feedback has shown to have strong influence in the computational properties and emergent behaviour of both classical and quantum systems \cite{Torres2002, Torres2007, Pantic2002, Mejias2009, Mejias2011,   Torres2013, Torres2022, TorresRomera2026}. In the following, we investigate the emergence of quantum phase transitions in this quantum brain model, thereby elucidating how synaptic feedback reshapes the phases and the associated transition lines of the LMG model. Note that, upon assuming $\lambda=\lambda_0 r(t)$, the quantum brain and LMG models coincide for $r=1$, a condition that can be attained in the limit $\tau_r \to 0$. Moreover, we perform a dynamical analysis that explicitly incorporates feedback-driven effects, with the aim of characterizing how such retroactive mechanisms nontrivially alter the system's temporal evolution.


\section{Results}

\begin{figure}[h!]
\begin{center}
\includegraphics[width=0.3333\linewidth]{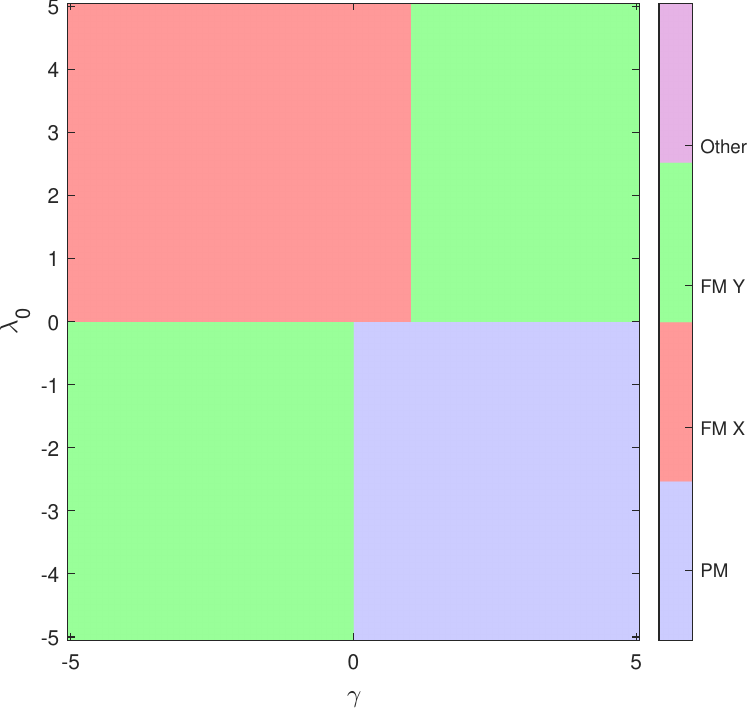}\hspace*{0.5cm}\includegraphics[width=0.33333\linewidth]{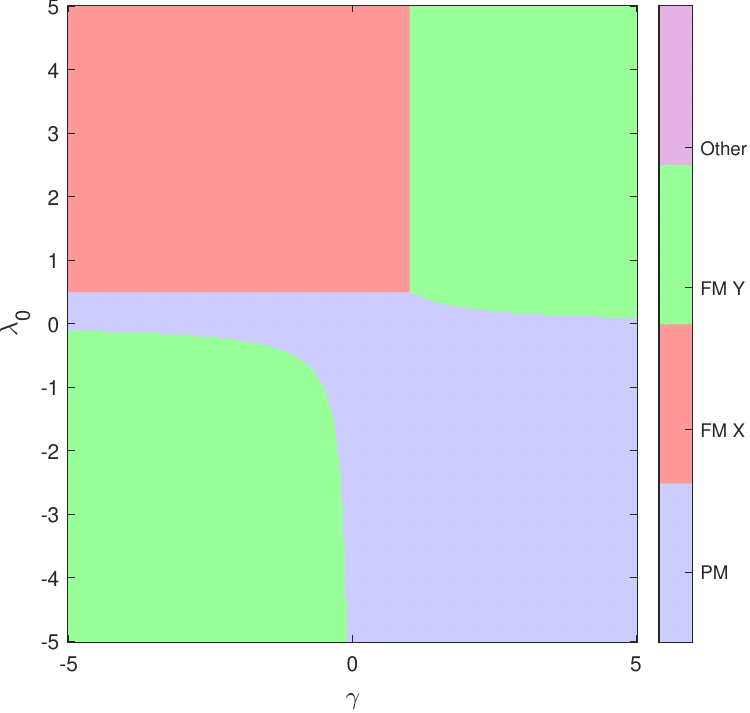}\hspace*{0.5cm}\includegraphics[width=0.33333\linewidth]{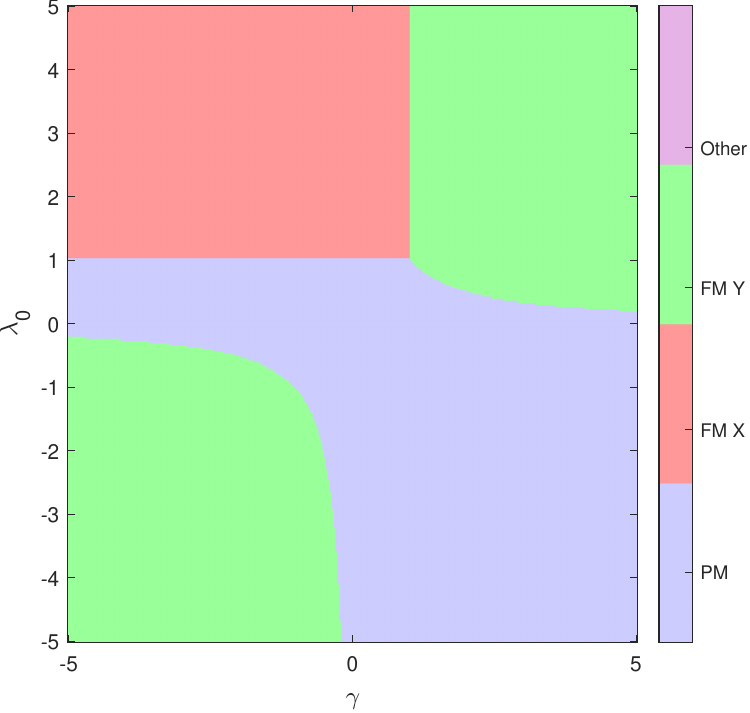}
\end{center}
\caption{Phase diagram of the LMG based brain model for $\tau_r\rightarrow 0$ (that is the standard LMG model),  $h=0$ (left), $h=0.5$ (center) and $h=1$ (right). Different colors corresponds to different phases with the code explained in the color bar on the right corresponding to FMX ($m_x\neq 0$), FMY ($m_y\neq0$) and PM ($m_x=m_y=0$ and $m_z\neq 0$)) }
\label{figura1}
\end{figure}

\begin{figure}
\begin{center}
\includegraphics[width=\linewidth]{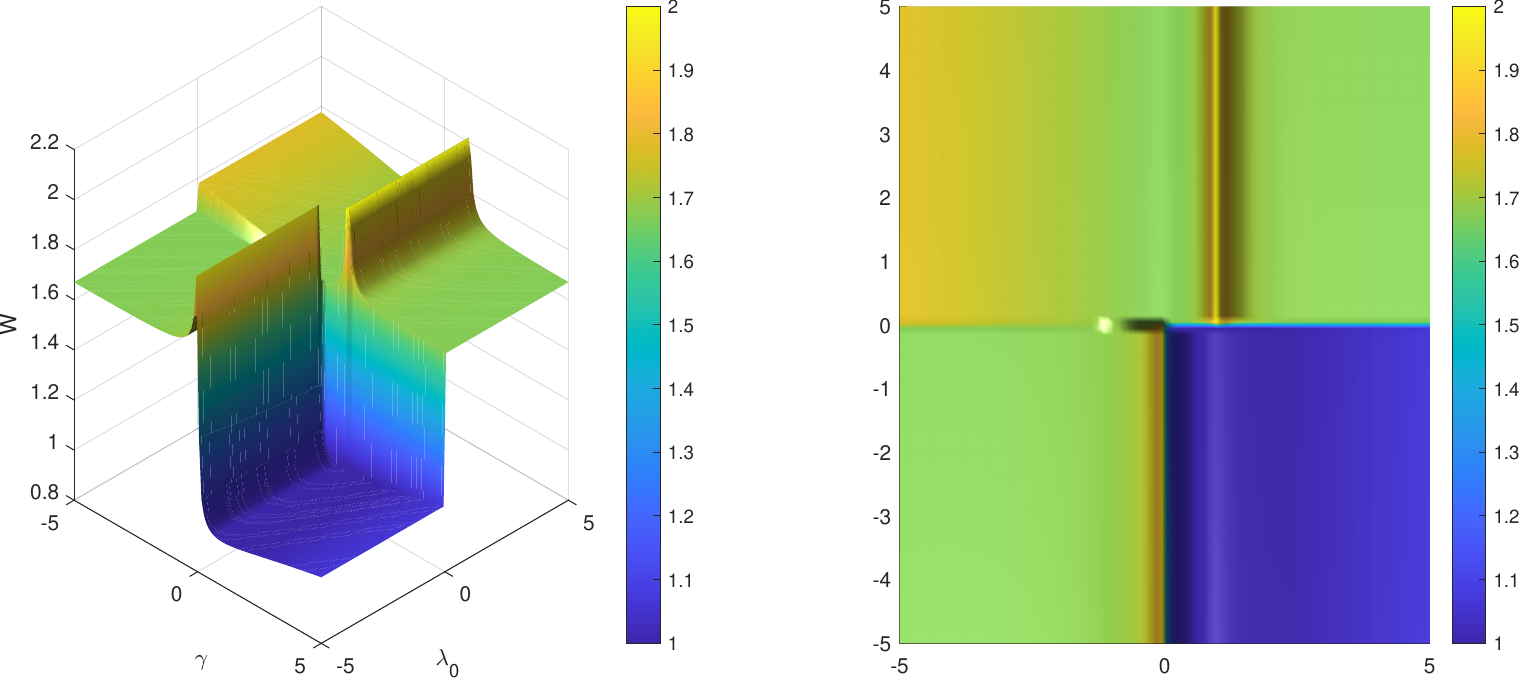}
\end{center}
\caption{Characterization of the quantum phase transitions of the LMG model ($\tau_r=0$) in terms of the Wehrl entropy, clearly depicting strong changes in the phase-space localitation associated to the quantum phase transitions illustrated in Fig.~\ref{figura1} for $h=0$.}
\label{fig:entropyh0dyntau0}
\end{figure}

\begin{figure}
\begin{center}
\includegraphics[width=\linewidth]{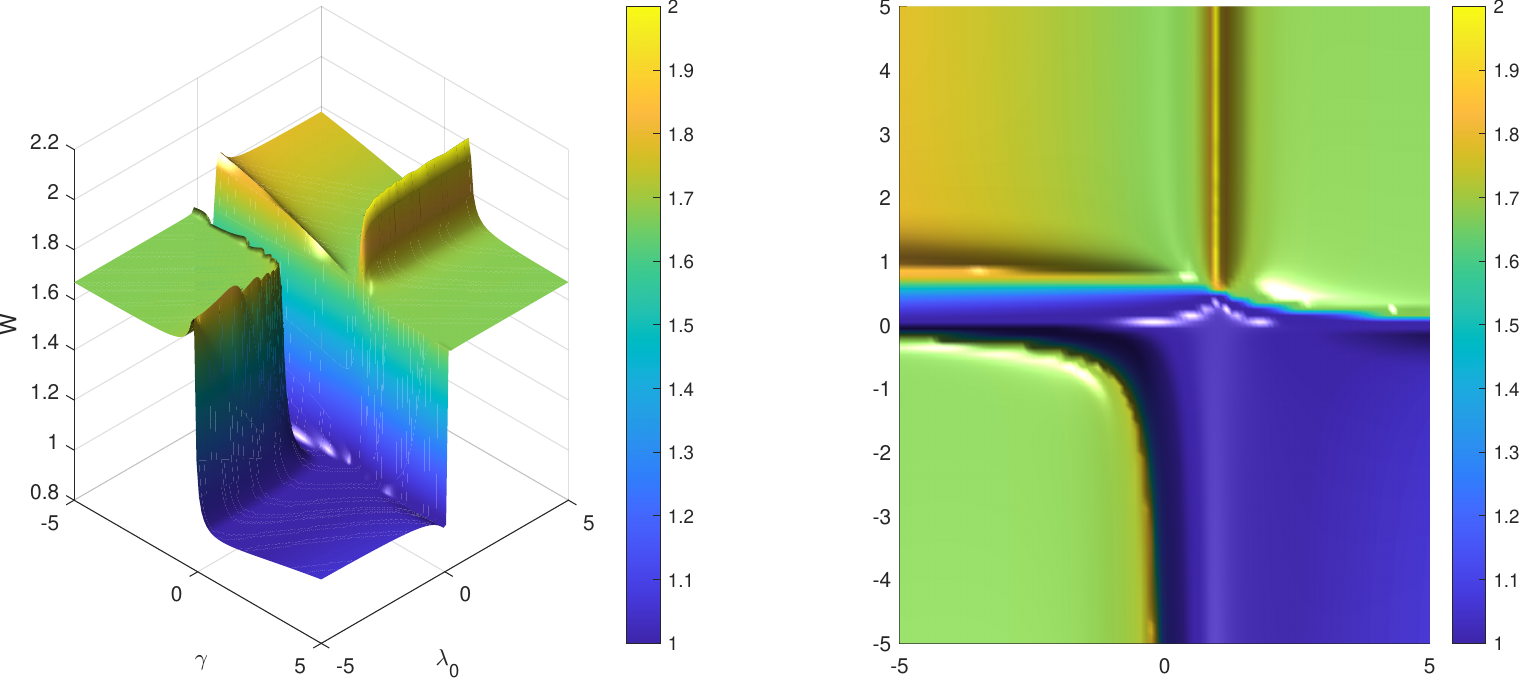}
\end{center}
\caption{Characterization of the quantum phase transitions of the LMG model ($\tau_r=0$) in terms of the Wehrl entropy, clearly depicting strong changes in the phase-space localitazion associated to the quantum phase transitions illustrated in Fig.~\ref{figura1} for $h=0.5$.}
\label{fig:entropyh05dyntau0}
\end{figure}

\begin{figure}
\begin{center}
\includegraphics[width=\linewidth]{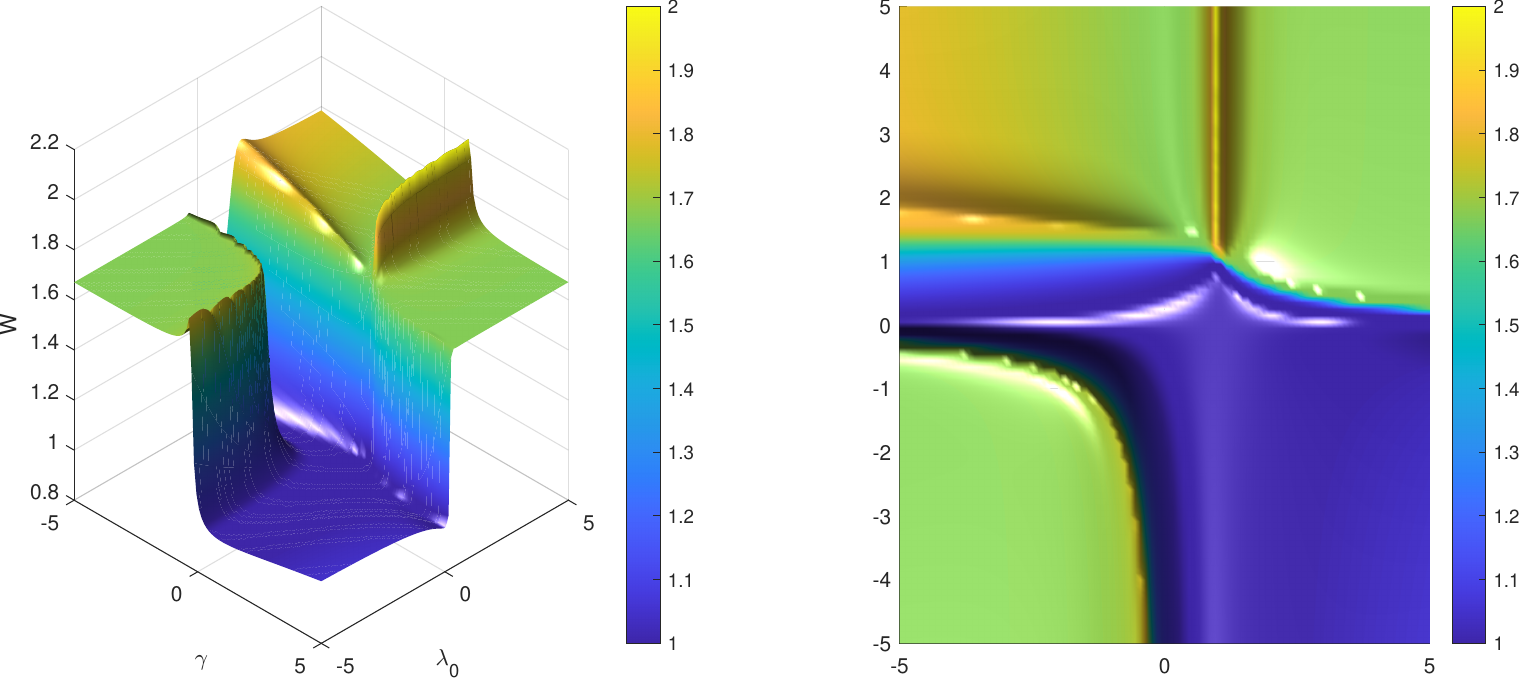}
\end{center}
\caption{Characterization of the quantum phase transitions of the LMG model ($\tau_r=0$) in terms of the Wehrl entropy, clearly depicting strong changes in the phase-space localitazion associated to the quantum phase transitions illustrated in Fig.~\ref{figura1} for $h=1$.}
\label{fig:entropyh1dyntau0}
\end{figure}

\subsection{Phase diagram of the LMG model}
We first provide a concise overview of the phase diagram of the LMG model and subsequently examine how its structure is altered when passing to the quantum brain model, maintaining throughout a phase--space perspective. As noted above, we may start from the system defined by Eq.~(8) in the limit $\tau_r \to 0$, which reduces to the LMG model. Figure~\ref{figura1} displays the corresponding phase diagrams in the $(\gamma,\lambda_0)$ parameter plane for three representative values of the effective field $h$, namely $0$, $0.5$, and $1$, with the phases encoded by different colors (see the figure caption). Each panel  exhibits a rich diagram, partitioned into regions associated with qualitatively distinct phases: the paramagnetic phase (PM), characterized by $m_x=m_y=0$ and $m_z\neq 0$, and the ferromagnetic phases FMX ($m_x\neq 0$) and FMY ($m_y\neq 0$). These diagrams are not merely descriptive; rather, they encode the quantum critical lines arising from the competition between the collective interaction (controlled by $\lambda_0$ and anisotropically modulated by $\gamma$) and the polarization term proportional to $h$. In particular, as $h$ increases, the PM phase expands, while FMX and FMY shrink slightly. For $h=0$ (left panel), the phase boundaries display a simple and highly symmetric geometry, yielding an almost ``quadrant--like'' partition, which indicates that the exchange of stability between collective configurations is governed rather directly by the relative signs and magnitudes of $\lambda_0$ and $\gamma$. For $h=0.5$ (central panel), the critical lines become curved, reflecting a reorganization of the energetic balance between phases: the effective field introduces a bias that reshapes the energy landscape and thereby affects the nature of the transitions induced by $(\lambda_0,\gamma)$. Finally, for $h=1$ (right panel), these deformations are further enhanced and the separatrices shift more substantially, underscoring that the term $hJ_z$ may amplify or suppress extended regions of parameter space, thereby modifying both the localization properties and the dominant character of the transitions.

A central point is that, upon varying $\lambda_0$ and $\gamma$, the system may undergo quantum phase transitions (QPTs) of first and second order. From a semiclassical standpoint, (i) first--order transitions are discontinuous: the global minimum changes abruptly between competing macroscopic configurations and typically arise when two local minima exchange stability, thereby producing a sharper separation between regions in parameter space; whereas (ii) second--order transitions are continuous and are associated with bifurcations in which the global minimum of the energy landscape evolves smoothly and the order parameter turns on gradually. To identify and characterize these phases and their transitions in a systematic manner from a phase--space viewpoint, we employ the Wehrl entropy associated with the Husimi function, which has proven to be a particularly effective diagnostic in this context and which are depicted in Figs. \ref{fig:entropyh0dyntau0}-\ref{fig:entropyh1dyntau0} for different values of $h$. Note that, as expected (see \cite{husimi1,husimi2,husimi3,Romera2014,Castanos2015}), the Wehrl entropy attains a maximum at the critical point of a first--order QPT and exhibits a step--function--like behavior at the critical point of a second--order QPT. In Fig.~\ref{fig:entropyh0dyntau0} ($h=0$), $W$ provides an especially transparent representation: extended domains with $W\simeq 1$ indicate strong Husimi localization and ground states well described by $SU(2)$ coherent states, in agreement with Lieb's asymptotic lower bound $W\to 1$; by contrast, regions with $W\simeq 1.7$ signal nonclassical ground states with two well--separated phase--space components (``cat''--like states), quantitatively consistent with $W\simeq 1+\ln 2\simeq 1.693$. In Fig.~\ref{fig:entropyh05dyntau0} ($h=0.5$), the same entropic signatures persist, but the interfaces between the low--entropy ($W\simeq 1$) and high--entropy ($W\simeq 1.7$) sectors become markedly more distorted and displaced, reflecting the increasing influence of the polarization term $hJ_z$, which reshapes the energy landscape and redistributes the regions in parameter space where the ground state remains coherent or develops a two--lobed phase--space structure. In Fig.~\ref{fig:entropyh1dyntau0} ($h=1$), this reorganization is even more pronounced: although the two characteristic plateaus remain essentially unchanged in magnitude, their corresponding regions in parameter space shift substantially and acquire a more intricate geometry; once again, $W\simeq 1$ identifies coherent regimes, whereas $W\simeq 1.7$ delineates extended nonclassical sectors, indicating the persistence of ``cat''--like structures with the additive contribution $\ln 2$. Overall, the evolution from Fig.~\ref{fig:entropyh0dyntau0} to Fig.~\ref{fig:entropyh1dyntau0} shows that increasing $h$ primarily relocalizes and reconfigures the coherent and cat--like domains rather than eliminating either of them.

\subsection{Phase diagram in the quantum brain model}
\begin{figure}[h!]
\begin{center}
\includegraphics[width=0.33333\linewidth]{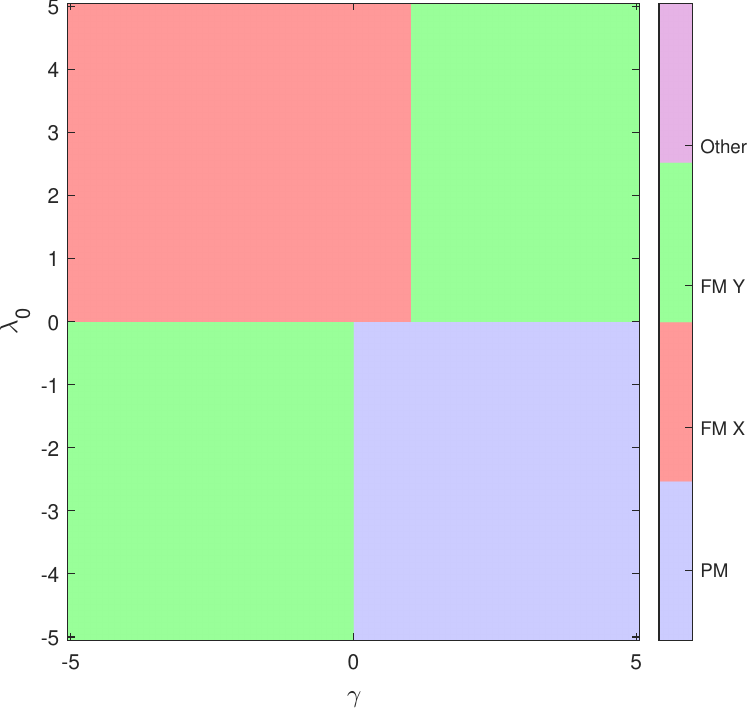}\hspace*{0.5cm}\includegraphics[width=0.3333\linewidth]{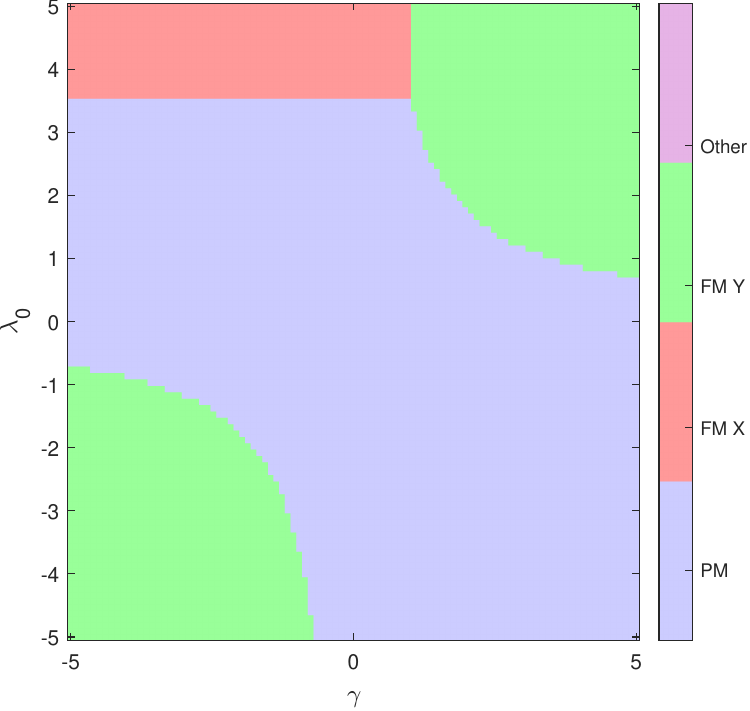}\hspace*{0.5cm}\includegraphics[width=0.3333\linewidth]{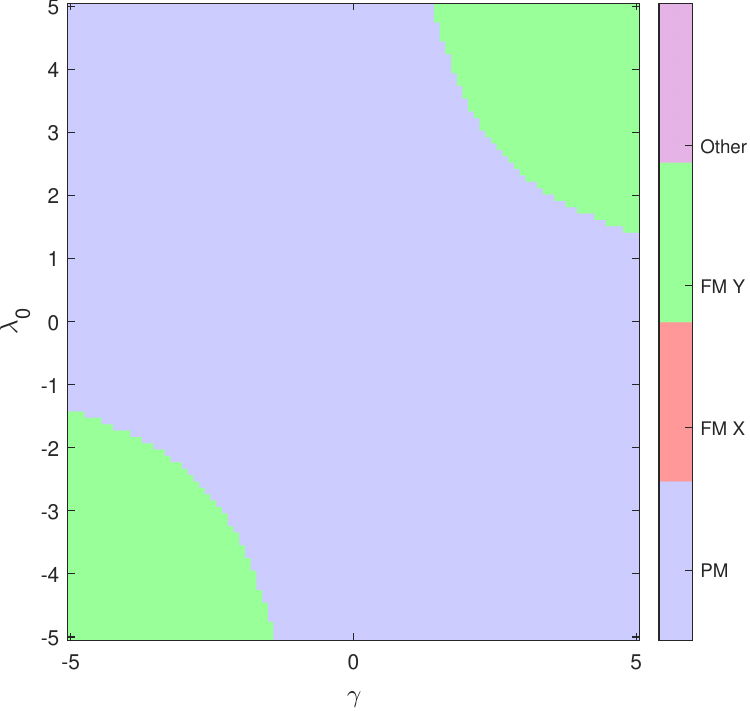}
\end{center}
\caption{Phase diagram of the LMG model with synaptic feedback (the so-called quantum brain model) for $h=0$ (left), $h=0.5$ (center) and $h=1$  (right). In all situations $\tau_r=10,$ $\mathcal{U}=0.5$ and $\tau_f=0.$}
\label{figurabrain}
\end{figure}

We can obtain the phase diagram of the quantum brain model by adopting the same adiabatic approach used above for the LMG model, namely by imposing steady-state conditions on the synaptic-feedback mechanism. In the stationary regime ($t\to\infty$), the system~(\ref{brainmodel}) yields
\begin{equation}
r_\infty=\frac{1}{1+\tau_rU_{\infty}E_\infty},
\qquad
U_{\infty}=\frac{\mathcal{U}(1+\tau_f E_\infty)}{1+\mathcal{U}\tau_f E_\infty},
\label{brainsemi}
\end{equation}
where $r_\infty$, $U_\infty$, and $E_\infty$ denote the asymptotic values of the dynamical variables entering (\ref{brainmodel}). Consequently, the phase diagram and the Husimi functions can be computed in direct analogy with the LMG case by performing the substitution $\lambda=\lambda_0 r_\infty$. In particular, in the limit $\tau_r\to 0$, the first expression in (\ref{brainsemi}) implies $r_\infty=1$, thereby recovering the standard LMG model phase portrait (see Fig.\ref{figura1}). More intriguing are the resulting phase diagrams under these steady-state conditions for $\tau_r>0$ (synaptic feedback is present in the quantum brain model) as it is summarized in Fig.~\ref{figurabrain} for three values of $h$ and for a feedback setting characterized by $\tau_r=10$, $\mathcal{U}=0.5$, and $\tau_f=0$ (for simplicity, we neglect facilitation). One finds that synaptic feedback modifies the phase diagram appreciably only for $h>0$. In this regime, it induces a substantial expansion of the paramagnetic (PM) phase, whose extent increases systematically with the synaptic feedback time scale $\tau_r$ (data not shown), at the expense of the ferromagnetic (FMX and FMY) phases. This behavior indicates that synaptic feedback enhances the stabilizing role of the external magnetic field, enlarging the parameter region in which the field-dominated paramagnetic state prevails.
This pronounced sensitivity to $h$ can be traced back to the specific manner in which feedback couples to the LMG dynamics. In particular, the feedback enters through $E(t)=(1+m_z)/2$, which depends explicitly on the longitudinal magnetization $m_z$. Since $m_z$ is strongly shaped by the Zeeman-like term $hJ_z$ in the Hamiltonian, the external field $h$ effectively modulates the feedback strength. As a result, for $h>0$ the field biases the system toward polarization along the $z$ direction, amplifies the impact of synaptic coupling, and stabilizes the PM phase over an enlarged region of parameter space.

\begin{figure}
\begin{center}
\includegraphics[width=\linewidth]{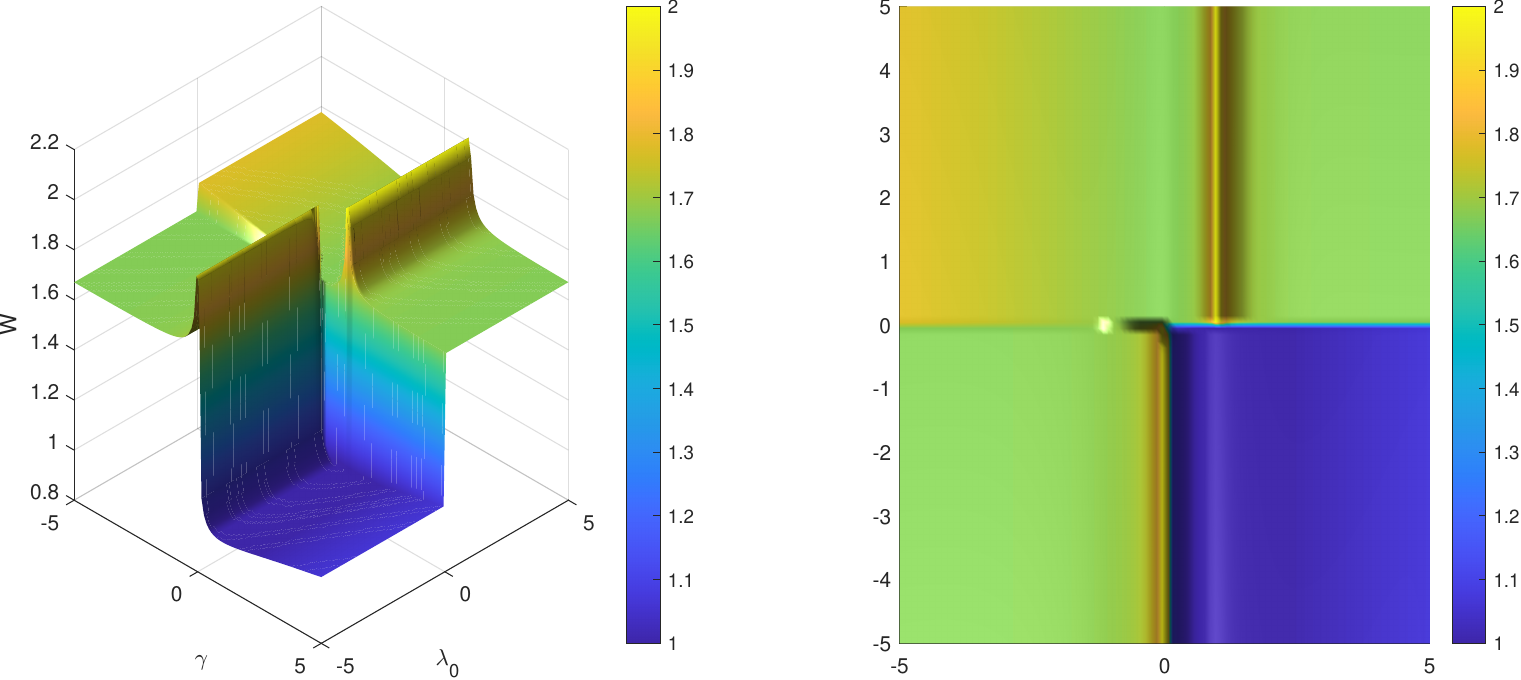}
\end{center}
\caption{Characterization of the quantum phase transitions of the Quantum brain model in terms of the Wehrl entropy, clearly depicting strong changes in the phase-space localization associated to the quantum phase transitions illustrated in Fig.~\ref{figurabrain} for $h=0$. Synaptic feedback parameters are $\tau_r=10,$ $\mathcal{U}=0.5$ and $\tau_f=0.$}
\label{fig:entropyh0dyntau10}
\end{figure}

\begin{figure}
\begin{center}
\includegraphics[width=\linewidth]{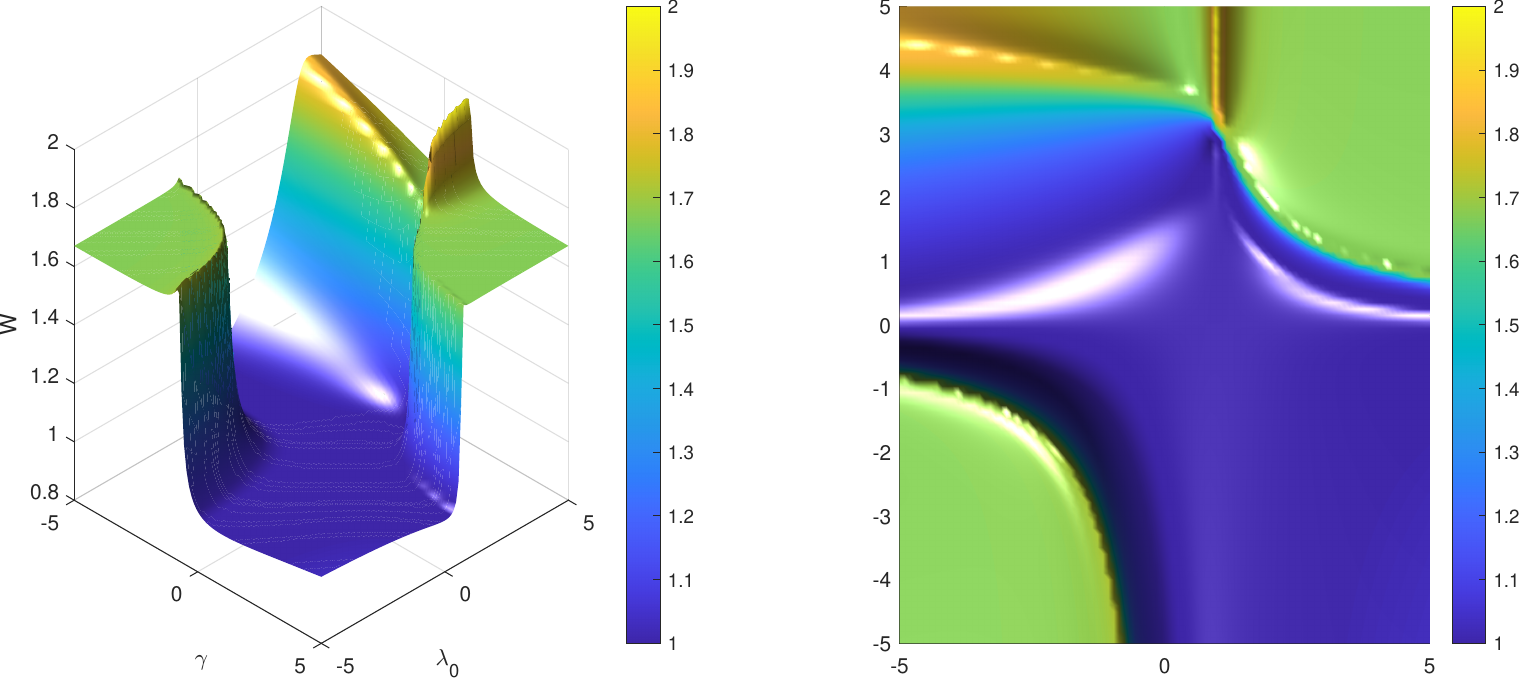}
\end{center}
\caption{Characterization of the quantum phase transitions of the Quantum brain model in terms of the Wehrl entropy, clearly depicting strong changes in the phase-space localization associated to the quantum phase transitions illustrated in Fig.~\ref{figurabrain} for $h=0.5$. Synaptic feedback parameters are $\tau_r=10,$ $\mathcal{U}=0.5$ and $\tau_f=0.$}
\label{fig:entropyh05dyntau10}
\end{figure}

\begin{figure}
\begin{center}
\includegraphics[width=\linewidth]{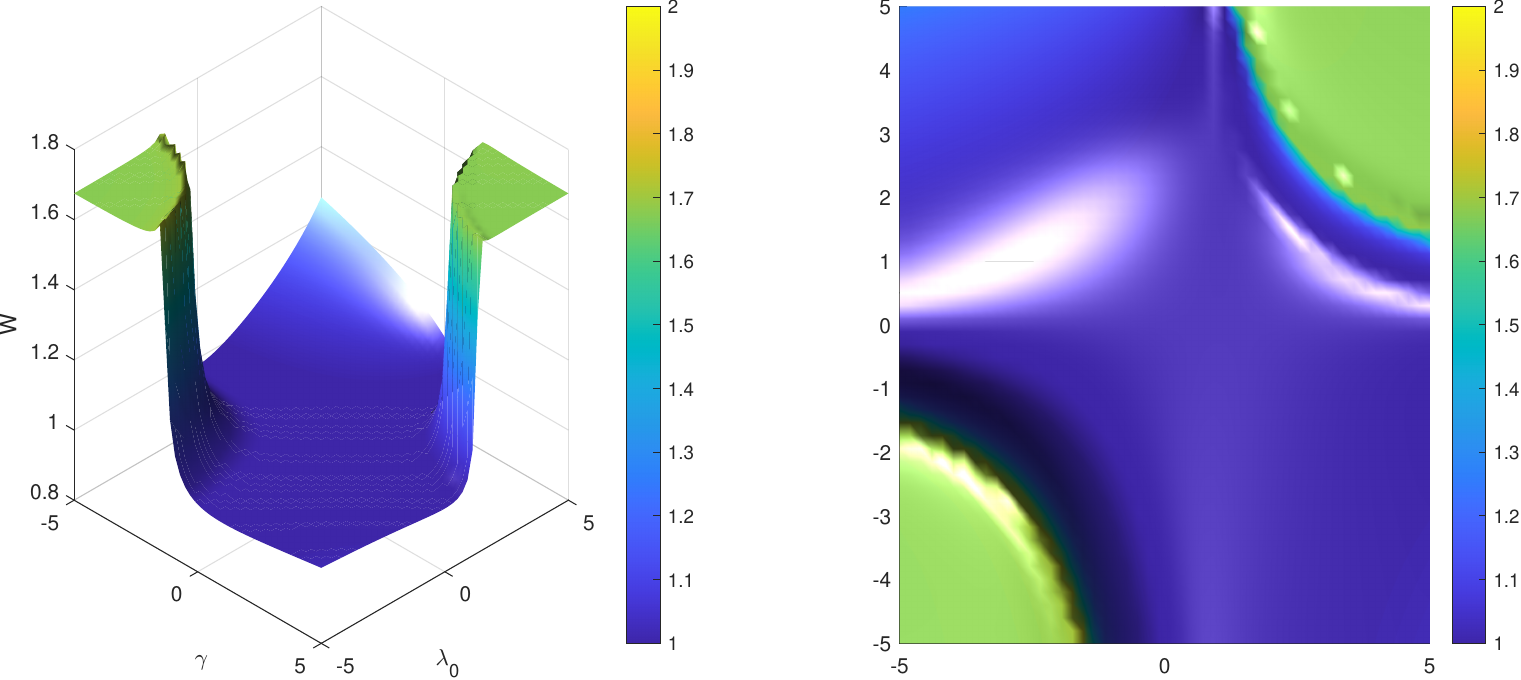}
\end{center}
\caption{Characterization of the quantum phase transitions of the Quantum brain model in terms of the Wehrl entropy, clearly depicting strong changes in the phase-space localization associated to the quantum phase transitions illustrated in Fig.~\ref{figurabrain} for $h=1$. Synaptic feedback parameters are $\tau_r=10,$ $\mathcal{U}=0.5$ and $\tau_f=0.$}
\label{fig:entropyh1dyntau10}
\end{figure}

This phenomenology can also be visualized, in close analogy with the LMG case, through the Wehrl entropy, here regarded as $W(\gamma,\lambda_0)$. In the $(\gamma,\lambda_0)$ plane, the Wehrl entropy exhibits the same qualitative patterns observed previously: it provides clear signatures of the phase transitions and captures the characteristic contrast between domains of higher and lower phase-space localization as the critical boundaries are crossed. This is illustrated in Fig.~\ref{fig:entropyh0dyntau10} for $h=0$, Fig.~\ref{fig:entropyh05dyntau10} for $h=0.5$, and Fig.~\ref{fig:entropyh1dyntau10} for $h=1$, respectively.



\subsection{Dynamical behaviour of the LMG brain model}
The static analysis developed in the previous sections rests on an adiabatic fixed-point approximation for the LMG-inspired quantum brain model. While this framework is well suited to identify and classify stationary phases, it does not capture the intrinsically dynamical phenomena that arise when synaptic feedback acts explicitly in time. In this regard, Ref.~\cite{TorresRomera2026} presents an exhaustive dynamical treatment of the model, incorporating genuinely non-adiabatic effects and granting access to time-dependent observables, thereby enabling the characterization of transient regimes, possible dynamical instabilities, and abrupt variations in the system's response upon crossing critical lines as the relevant control parameters are varied. Nevertheless, with the aim of faithfully characterizing the global behavior of the system, it is necessary to complement the static picture with a dynamical analysis that explicitly incorporates the effects induced by synaptic feedback.

To derive a mean-field description, we consider a system of $N$ spins and adopt the standard scaling $j=N/2$. We recompute the mean energy per particle for the brain model following the same procedure described in Sec.\ref{sec:semi}. In particular, evaluating the effective Hamiltonian on spin-coherent states yields the energy surface
\[
\mathcal{E}(\theta,\phi)=
-\frac{\lambda}{4}\sin^2\theta\Big(\cos^2\phi+\gamma\sin^2\phi\Big)
-\frac{h}{2}\cos\theta,
\]
where, in the brain-model setting, the coupling should be understood as $\lambda=\lambda_0\,r(t)$ (or $\lambda=\lambda_0\,r_\infty$ in the stationary regime), as appropriate.
Taking $q=\phi$ and $p=J_z=j\cos\theta$ as canonical variables, the mean-field dynamics follows from the Poisson bracket, leading to Hamilton's equations for $\theta$ and $\phi$, namely $\dot\theta=\{\theta,\mathcal{E}\}$ and $\dot\phi=\{\phi,\mathcal{E}\}$. 
So, the extension to the LMG-based brain model is immediate: synaptic feedback induces a time-dependent modulation of the collective coupling through $r(t)$, so that the dynamics of $(\theta,\phi)$ becomes coupled to the evolution of $(r,U)$ dictated by synaptic plasticity. The resulting system can be written as
\begin{equation}
\dot\theta=\frac{\lambda_0 r}{2}(1-\gamma)\sin\theta\sin(2\phi),
\qquad
\dot\phi=\lambda_0 r\cos\theta\big(\cos^2\phi+\gamma\sin^2\phi\big)-h,
\label{dyneq1}
\end{equation}
\begin{equation}
\dot r=\frac{1-r}{\tau_r}-Ur\,\frac{1+\cos\theta}{2},
\qquad
\dot U=\frac{\mathcal{U}-U}{\tau_f}+\mathcal{U}(1-U)\,\frac{1+\cos\theta}{2}.
\label{dyneq}
\end{equation}
These expressions constitute the mean-field dynamical equations for coherent-state evolutions of the LMG-based brain model and provide a natural benchmark against the full quantum dynamics.

\begin{figure}
\begin{center}
\includegraphics[width=12cm]{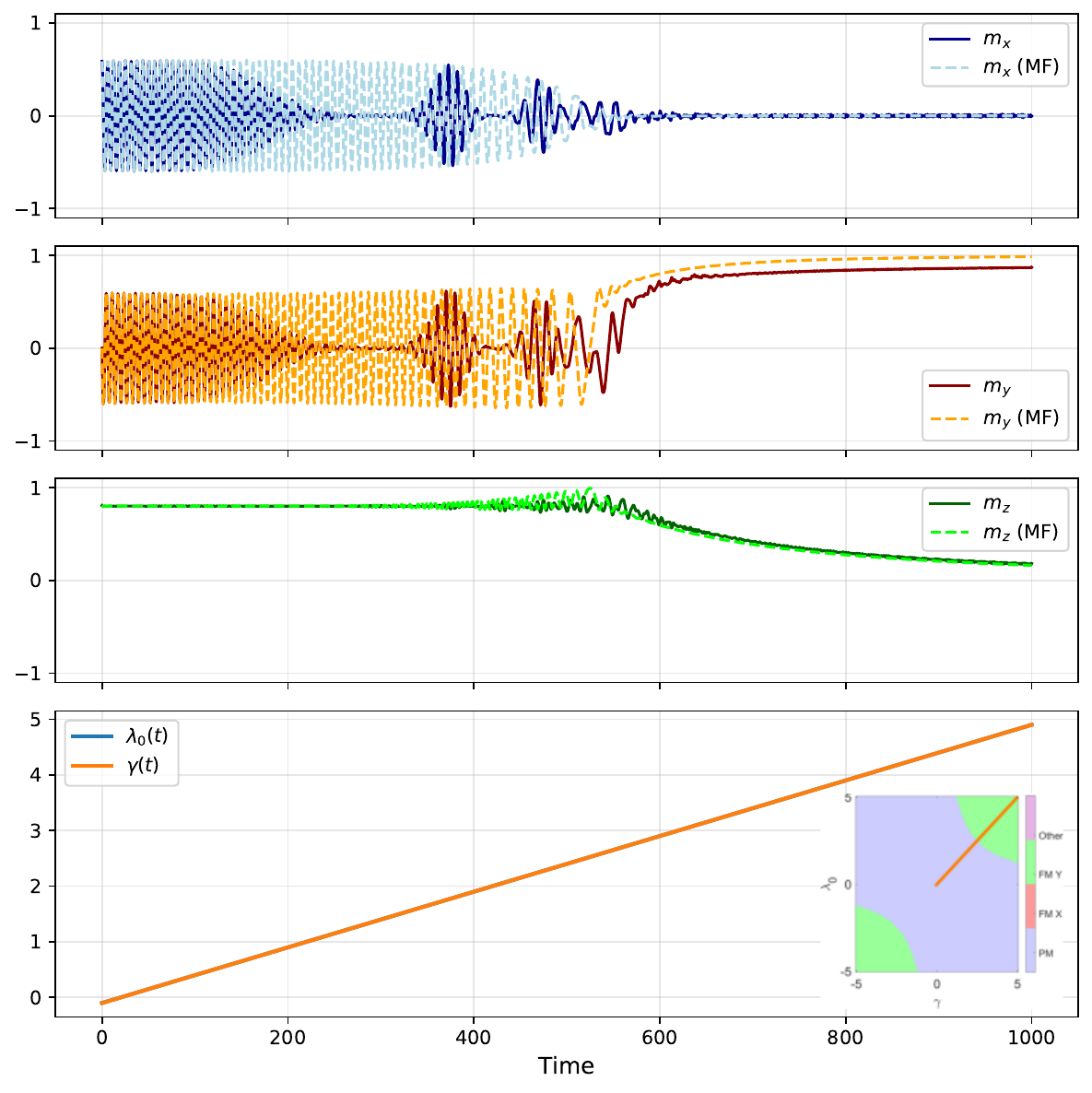}
\end{center}
\caption{The figure shows the evolution of the order paraameters $m_\alpha$ for the LMG brain model starting from a coherent state (see main text) along the relevant trajectory $\lambda_0=\gamma$ with paramters linearly increasing in time from $-0.1$  to 5 (see bottom panel) for $\tau_r=10,$ $\tau_f=0,$ $\mathcal{U}=0.5$ and $h=1$ passing from PM to FMY phases (see also figure \ref{figurabrain} and figure \ref{fig:entropyh1dyntau10}). From top to third panel the figure shows the quantum evolution of magnetizations $m_\alpha$ for $N=20$ (solid lines) and their evolution through the mean field dynamical equations (\ref{dyneq1}--\ref{dyneq}). The figure shows an almost perfect agreement between quantum evolution and mean field equations derived for the quantum brain model. }
\label{figdyn}
\end{figure}

It is worth noting to say that from the dynamics in Eqs (\ref{dyneq1}-\ref{dyneq}) one can easily monitor $m_\alpha(t)$ using (\ref{eq:Jsymbols}). Thus, in Fig.~\ref{figdyn} we report the semiclassical simulation of the system through the curves $m_x(t)$, $m_y(t)$, and $m_z(t)$ which describe the time evolution of the collective magnetization treated as a unit classical vector (parametrized by $\theta(t)$ and $\phi(t)$), i.e., the precession of the ``macroscopic spin'' under the field $h$ and the model's effective coupling (with parameters ramped linearly in time). Within this mean-field framework,
\[
m_x=\sin\theta\cos\phi,\qquad m_y=\sin\theta\sin\phi,\qquad m_z=\cos\theta,
\]
so that the trajectory corresponds directly to the motion of the Bloch vector $\vec m(t)$ with $|\vec m(t)|=1$. In the quantum simulation, the analogous quantities $m_\alpha(t)=\langle J_\alpha\rangle/(N/2)$ ($\alpha=x,y,z$) quantify the mean orientation of the collective spin from the state $\psi(t)$ evolving unitarily under a time-dependent Hamiltonian. We have focus in a trajectory $\lambda_0=\gamma$ in the ($\gamma,\lambda_0$) plane that starting from the center of the paramagnetic phase (PM) cross a phase transition and finish in the ferromagnetic phase FMY (see the red line in the inset of the bottom panel of the Fig. \ref{figdyn}). In contrast to the semiclassical case, these observables incorporate the build-up of genuine quantum correlations, so that the ``mean'' vector may contract even though the evolution remains strictly unitary. This behavior is visible in Fig.~\ref{figdyn}: while both descriptions agree initially, they gradually diverge as correlations develop at the phase transition, consistently with the phase diagram discussed in the previous section.
More precisely, although the longitudinal component $m_z(t)$ is reproduced with remarkable accuracy over essentially the entire protocol, the transverse components $m_x(t)$ and, more noticeably, $m_y(t)$ exhibit systematic deviations. In this case, the quantum curves display an earlier and stronger contraction of the oscillation envelope than their mean-field counterparts.
This behavior is consistent with the progressive build-up of genuine quantum correlations, which induces an effective dephasing of the collective motion (without implying any dissipative loss), an effect that is absent at the mean-field level. 
Also for the transversal magnetizations, the time traces clearly delineate a  transient around the central part of the evolution, characterized by a burst of enhanced oscillatory activity followed by a rapid crossover to a near-stationary regime. This regime signals a pronounced reorganization of the effective energy landscape, as expected when the system is driven across stability boundaries (e.g., separatrices) or through critical regions of the phase diagram. 
Finally, the asymptotic behavior provides a direct dynamical fingerprint of the emergent ordering: $m_x(t)$ is progressively suppressed, whereas $m_y(t)$ grows and saturates, indicating that the trajectory ultimately selects an FMY-like ordered configuration. Quantitatively, the mean-field dynamics tends to \emph{overestimate} the final degree of ordering (with $m_y$ approaching a larger saturation value than in the quantum simulation), again reflecting the fact that quantum fluctuations and phase-space delocalization reduce the macroscopic order parameter relative to the purely coherent mean-field prediction.

\section{Conclusions}
In this work we investigate the emergence of phase transitions in a recently proposed quantum brain model built upon the Lipkin--Meshkov--Glick framework, in which a biologically motivated synaptic feedback nonlinearly and dynamically modulates the collective interaction. We show that this state-dependent coupling substantially reshapes the phase structure across parameter space. Thus, relative to the feedback-free case, the paramagnetic phase expands at the expense of the ferromagnetic phases, demonstrating that feedback induces a genuine macroscopic reorganization rather than a marginal correction. This effect becomes particularly pronounced in the presence of a longitudinal external field, since the feedback is activated through a quantity controlled by the longitudinal magnetization. Consequently, the field biases the system toward longitudinal polarization and amplifies the action of the synaptic feedback, which may be interpreted as a state-dependent renormalization of the effective coupling and a concomitant displacement of the critical boundaries.

The ensuing transitions are characterized from a phase-space perspective using the Husimi distribution of the ground state and the Wehrl entropy, which we confirm as a robust diagnostic of qualitative changes in localization. In particular, these quantities allow one to clearly identify highly localized regions and more delocalized, nonclassical regimes, while faithfully tracking the feedback-induced deformation of the phase boundaries on the Bloch sphere.

To capture the intrinsically temporal phenomenology associated with the evolution of the feedback, we perform an explicit dynamical analysis. To this end, we derive mean-field equations for the angular variables describing the collective-spin orientation, self-consistently coupled to the dynamics of the synaptic variables. For the protocols considered, this framework reproduces the time evolution of collective observables with high fidelity relative to the full quantum dynamics, thereby providing a useful benchmark for understanding the system's global behavior.

Finally, notwithstanding the overall agreement between the two descriptions, the quantum dynamics displays signatures consistent with the build-up of genuinely quantum correlations, manifested as an effective dephasing of the collective motion and a stronger attenuation of transverse components than predicted at the mean-field level. Moreover, upon traversing critical regions, the temporal traces exhibit nontrivial transients -- including episodes of enhanced oscillatory activity followed by rapid crossovers to near-stationary regimes -- reflecting a feedback-induced reorganization of the effective energy landscape. Along these lines, it will be particularly interesting to undertake a systematic investigation of decoherence and environmental coupling mechanisms in this quantum brain model, in order to assess their impact on phase stability, transient dynamics, and the formation of nonclassical correlations.

\section{Acknowledgments}
E.R. acknowledges support from PAIDI group FQM-420 of the University of Granada.  J.J.T. acknowledges support by Grant No. PID2023-149174NB-I00 financed by the Spanish Ministry and Agencia Estatal de Investigación MICIU/AEI/10.13039/501100011033 and ERDF funds (European Union).  

\section{Author contributions}
E.R. and J.J.T. equally contributed to the conception,  design, and research of the methods presented in this article. All authors equally contributed to the writing of the manuscript.
\section{Competing interests}
The authors declare no competing interests.
\bibliographystyle{plain} 
\bibliography{references} 

@article{Romera2014,
doi = {10.1088/0031-8949/89/9/095103},
url = {https://doi.org/10.1088/0031-8949/89/9/095103},
year = {2014},
month = {aug},
publisher = {IOP Publishing},
volume = {89},
number = {9},
pages = {095103},
author = {Romera, Elvira and Calixto, Manuel and Castaños, Octavio},
title = {Phase space analysis of first-, second- and third-order quantum phase transitions in the Lipkin–Meshkov–Glick model},
journal = {Physica Scripta}
}

@article{Castanos2005a,
  title = {Phase transitions and accidental degeneracy in nonlinear spin systems},
  author = {Casta\~nos, Octavio and L\'opez-Pe\~na, Ram\'on and Hirsch, Jorge G. and L\'opez-Moreno, Enrique},
  journal = {Phys. Rev. B},
  volume = {72},
  issue = {1},
  pages = {012406},
  numpages = {4},
  year = {2005},
  month = {Jul},
  publisher = {American Physical Society},
  doi = {10.1103/PhysRevB.72.012406},
  url = {https://link.aps.org/doi/10.1103/PhysRevB.72.012406}
}

@article{Castanos2015,
  title = {Identifying the order of a quantum phase transition by means of Wehrl entropy in phase space},
  author = {Casta\~nos, Octavio and Calixto, Manuel and P\'erez-Bernal, Francisco and Romera, Elvira},
  journal = {Phys. Rev. E},
  volume = {92},
  issue = {5},
  pages = {052106},
  numpages = {7},
  year = {2015},
  month = {Nov},
  publisher = {American Physical Society},
  doi = {10.1103/PhysRevE.92.052106},
  url = {https://link.aps.org/doi/10.1103/PhysRevE.92.052106}
}

@article{GnutzmannZyczkowski2001,
doi = {10.1088/0305-4470/34/47/317},
url = {https://doi.org/10.1088/0305-4470/34/47/317},
year = {2001},
month = {nov},
publisher = {},
volume = {34},
number = {47},
pages = {10123},
author = {Sven Gnutzmann and Karol Zyczkowski},
title = {Rényi-Wehrl entropies as measures of
localization in phase space},
journal = {Journal of Physics A: Mathematical and General}
}

@book{BengtssonZyczkowski2017,
  title     = {Geometry of Quantum States: An Introduction to Quantum Entanglement},
  author    = {Bengtsson, Ingemar and {\.Z}yczkowski, Karol},
  edition   = {2nd},
  year      = {2017},
  publisher = {Cambridge University Press},
  address   = {Cambridge, UK},
  isbn      = {9781107026254},
  doi       = {10.1017/9781139207010}
}

@article{Arecchi1972,
  title = {Atomic Coherent States in Quantum Optics},
  author = {Arecchi, F. T. and Courtens, Eric and Gilmore, Robert and Thomas, Harry},
  journal = {Phys. Rev. A},
  volume = {6},
  issue = {6},
  pages = {2211--2237},
  numpages = {0},
  year = {1972},
  month = {Dec},
  publisher = {American Physical Society},
  doi = {10.1103/PhysRevA.6.2211},
  url = {https://link.aps.org/doi/10.1103/PhysRevA.6.2211}
}

@article{TorresRomera2026,
  title   = {Dynamic Synaptic Modulation of LMG Qubits populations in a Bio-Inspired Quantum Brain},
  author  = {Torres, J.J. and Romera, E.},
  journal = {arXiv:2602.16003 [quant-ph]},
  year    = {2026},
}

@article{Romera2026HusimiLMG,
  author  = {Romera, E. and Calixto, M. and Casta{\~n}os, O.},
  title   = {Quantum phase transitions in the Lipkin–Meshkov–Glick model via mutual information within Husimi phase space representation},
  journal = {Physical Review E},
  year    = {2026},
  note    = {Accepted manuscript},
  doi     = {10.1103/r89j-9mjd}
}

@book{PerelomovBook,
  author    = {Perelomov, A. M.},
  title     = {Generalized Coherent States and Their Applications},
  publisher = {Springer},
  address   = {Berlin},
  year      = {1986}
}

@book{GilmoreBook,
  author    = {Gilmore, R.},
  title     = {Lie Groups, Lie Algebras, and Some of Their Applications},
  publisher = {Dover},
  address   = {New York},
  year      = {2006}
}

@article{Wehrl1978,
  author  = {Wehrl, A.},
  title   = {General properties of entropy},
  journal = {Reviews of Modern Physics},
  volume  = {50},
  pages   = {221--260},
  year    = {1978},
  doi     = {10.1103/RevModPhys.50.221}
}

@article{LiebSolovej2014,
  author  = {Lieb, Elliott H. and Solovej, Jan Philip},
  title   = {Proof of an entropy conjecture for Bloch coherent spin states and its generalizations},
  journal = {Acta Mathematica},
  volume  = {212},
  pages   = {379--398},
  year    = {2014},
  doi     = {10.1007/s11511-014-01},
}

@article{PhysRevA.108.023722,
  title   = {Spin squeezing with arbitrary quadratic collective-spin interactions},
  author  = {Hu, Zhiyao and Li, Qixian and Zhang, Xuanchen and Huang, Long-Gang and Zhang, He-bin and Liu, Yong-Chun},
  journal = {Phys. Rev. A},
  volume  = {108},
  number  = {2},
  pages   = {023722},
  year    = {2023},
  month   = aug,
  publisher = {American Physical Society},
  doi     = {10.1103/PhysRevA.108.023722},
  url     = {https://link.aps.org/doi/10.1103/PhysRevA.108.023722}
}

@article{Torres2022,
   author = {J J Torres and D Manzano},
   doi = {10.1088/1367-2630/ac7aaa},
   issn = {1367-2630},
   issue = {7},
   journal = {New Journal of Physics},
   month = {7},
   pages = {073007},
   title = {A model of interacting quantum neurons with a dynamic synapse},
   volume = {24},
   url = {https://iopscience.iop.org/article/10.1088/1367-2630/ac7aaa},
   year = {2022}
}

@article{Kristensen2021SpikingQuantumNeuron,
  title        = {An artificial spiking quantum neuron},
  author       = {Kristensen, Lars Bojer and Degroote, Matthias and Wittek, Peter and Aspuru-Guzik, Al{\'a}n and Zinner, Nikolaj Thomas},
  journal      = {npj Quantum Information},
  volume       = {7},
  pages        = {59},
  year         = {2021},
  doi          = {10.1038/s41534-021-00394-w},
  url          = {https://doi.org/10.1038/s41534-021-00394-w}
}

@article{Tacchino2019ArtificialNeuron,
  title        = {An artificial neuron implemented on an actual quantum processor},
  author       = {Tacchino, Francesco and Macchiavello, Chiara and Gerace, Dario and Bajoni, Daniele},
  journal      = {npj Quantum Information},
  volume       = {5},
  pages        = {26},
  year         = {2019},
  doi          = {10.1038/s41534-019-0140-4},
  url          = {https://doi.org/10.1038/s41534-019-0140-4}
}

@article{Cao2017QuantumNeuron,
  title        = {Quantum Neuron: An Elementary Building Block for Machine Learning on Quantum Computers},
  author       = {Cao, Yudong and Guerreschi, Gian Giacomo and Aspuru-Guzik, Al{\'a}n},
  journal      = {arXiv preprint arXiv:1711.11240},
  year         = {2017},
  url          = {https://arxiv.org/abs/1711.11240}
}

@article{Pechal2021QuantumPerceptron,
  title        = {Direct implementation of a perceptron in superconducting circuit quantum hardware},
  author       = {Pechal, Miroslav and Roy, Finn and Wilkinson, Samuel A. and Salis, Gian and Werninghaus, Max and Hartmann, Michael J. and Filipp, Stefan},
  journal      = {arXiv preprint arXiv:2111.12669},
  year         = {2021},
  url          = {https://arxiv.org/abs/2111.12669}
}

@article{Torres2013,
  author    = {Joaquín J. Torres and Hilbert J. Kappen},
  title     = {Emerging phenomena in neural networks with dynamic synapses and their computational implications},
  journal   = {Frontiers in Computational Neuroscience},
  volume    = {7},
  pages     = {30},
  year      = {2013},
  doi       = {10.3389/fncom.2013.00030},
}

@article{Torres2002,
  author    = {Joaquín J. Torres and Lovorca Pantic and Hilbert J. Kappen},
  title     = {Storage capacity of attractor neural networks with depressing synapses},
  journal   = {Physical Review E},
  volume    = {66},
  pages     = {061910},
  year      = {2002},
  doi       = {10.1103/PhysRevE.66.061910},
}

@article{Mejias2009,
  author    = {Jorge F. Mejías and Joaquín J. Torres},
  title     = {Maximum memory capacity on neural networks with short-term synaptic depression and facilitation},
  journal   = {Neural Computation},
  volume    = {21},
  pages     = {851--871},
  year      = {2009},
  doi       = {10.1162/neco.2008.03-07-479},
}

@article{Pantic2002,
  author    = {Lovorka Pantic and Joaquín J. Torres and Hilbert J. Kappen and Stan C. A. M. Gielen},
  title     = {Associative memory with dynamic synapses},
  journal   = {Neural Computation},
  volume    = {14},
  pages     = {2903--2923},
  year      = {2002},
  doi       = {10.1162/089976602760805359},
}

@article{Torres2007,
  author    = {Joaquín J. Torres and Jesus M. Cortes and Joaquín Marro and Hilbert J. Kappen},
  title     = {Competition between synaptic depression and facilitation in attractor neural networks},
  journal   = {Neural Computation},
  volume    = {19},
  pages     = {2739--2755},
  year      = {2007},
  doi       = {10.1162/neco.2007.19.10.2739},
}

@article{Menesse2024,
  author    = {Gustavo Menesse and Joaqu{\'\i}n J. Torres},
  title     = {Information dynamics of in silico {EEG} Brain Waves: Insights into oscillations and functions},
  journal   = {PLoS Computational Biology},
  volume    = {20},
  number    = {9},
  pages     = {e1012369},
  year      = {2024},
  doi       = {10.1371/journal.pcbi.1012369},
}

@article{Pretel2021,
  author    = {Jorge Pretel and Joaqu{\'\i}n J. Torres and Joaqu{\'\i}n Marro},
  title     = {{EEGs} Disclose Significant Brain Activity Correlated with Synaptic Fickleness},
  journal   = {Biology},
  volume    = {10},
  number    = {7},
  pages     = {647},
  year      = {2021},
  doi       = {10.3390/biology10070647},
}

@article{Mejias2011,
  author    = {Jorge F. Mejías and Joaquín J. Torres},
  title     = {Emergence of resonances in neural systems: the interplay between adaptive threshold and short-term synaptic plasticity},
  journal   = {PLoS ONE},
  volume    = {6},
  pages     = {e17255},
  year      = {2011},
  doi       = {10.1371/journal.pone.0017255},
}

@article{njp,
  author  = {Aulbach, C. and Wobst, A. and Ingold, G.-L. and H{\"a}nggi, P. and Varga, I.},
  title   = {Phase-space visualization of a metal--insulator transition},
  journal = {New Journal of Physics},
  volume  = {6},
  pages   = {70},
  year    = {2004},
  doi     = {10.1088/1367-2630/6/1/070}
}

@article{leboeuf,
  author  = {Leboeuf, P. and Voros, A.},
  title   = {Chaos-revealing multiplicative representation of quantum eigenstates},
  journal = {Journal of Physics A: Mathematical and General},
  volume  = {23},
  number  = {10},
  pages   = {1765--1774},
  year    = {1990},
  doi     = {10.1088/0305-4470/23/10/017}
}

@article{husimi1,
  author  = {Romera, E. and del Real, R. and Calixto, M.},
  title   = {Husimi distribution and phase-space analysis of a Dicke-model quantum phase transition},
  journal = {Physical Review A},
  volume  = {85},
  pages   = {053831},
  year    = {2012},
  doi     = {10.1103/PhysRevA.85.053831}
}

@article{husimi2,
  author  = {Calixto, M. and del Real, R. and Romera, E.},
  title   = {Husimi distribution and phase-space analysis of a vibron-model quantum phase transition},
  journal = {Physical Review A},
  volume  = {86},
  pages   = {032508},
  year    = {2012},
  doi     = {10.1103/PhysRevA.86.032508}
}

@article{husimi3,
  author  = {del Real, R. and Calixto, M. and Romera, E.},
  title   = {The Husimi distribution, the Wehrl entropy and the superradiant phase in spin-boson interactions},
  journal = {Physica Scripta},
  volume  = {T153},
  pages   = {014016},
  year    = {2013},
  doi     = {10.1088/0031-8949/2013/T153/014016}
}

@article{Tsodyks1998,
  author    = {Misha Tsodyks and Klaus Pawelzik and Henry Markram},
  title     = {Neural networks with dynamic synapses},
  journal   = {Neural Computation},
  volume    = {10},
  number    = {4},
  pages     = {821--835},
  year      = {1998},
  doi       = {10.1162/089976698300017502},
}

@article{Menesse2025AstrocyteControl,
  title        = {Astrocyte-Mediated Higher-Order Control of Synaptic Plasticity},
  author       = {Menesse, Gustavo and Millán, Ana P. and Torres, Joaquín J.},
  journal      = {arXiv preprint arXiv:2507.07693 [q-bio.NC]},
  year         = {2025},
  url          = {https://arxiv.org/abs/2507.07693}
}

@book{Amit2012ModelingBrainFunction,
  author       = {Amit, Daniel J.},
  title        = {Modeling Brain Function},
  publisher    = {Cambridge University Press},
  address      = {Cambridge},
  year         = {2012},
  isbn         = {9780521290301}
}

@inproceedings{Wiebe2016QuantumPerceptron,
  title        = {Quantum perceptron models},
  author       = {Wiebe, Nathan and Kapoor, Ashish and Svore, Krysta M.},
  booktitle    = {Proceedings of the 30th Conference on Neural Information Processing Systems (NeurIPS 2016)},
  year         = {2016},
  url          = {https://papers.nips.cc/paper/2016/hash/2f7123fd2f22f4b0a32b47b021b35f5b-Abstract.html}
}

@article{Rotondo2018,
  author    = {Rotondo, Pietro and Marcuzzi, Matteo and Garrahan, Juan P. and Lesanovsky, Igor and Müller, Markus},
  title     = {Open quantum generalisation of Hopfield neural networks},
  journal   = {Journal of Physics A: Mathematical and Theoretical},
  volume    = {51},
  number    = {11},
  pages     = {115301},
  year      = {2018},
  doi       = {10.1088/1751-8121/aaabcb},
  url       = {https://iopscience.iop.org/article/10.1088/1751-8121/aaabcb},
  eprint    = {1701.01727},
  archivePrefix = {arXiv},
  primaryClass  = {cond-mat.dis-nn}
}

@article{Torres2024,
   author = {Joaquín J Torres and Daniel Manzano},
   doi = {10.1088/1367-2630/ad5e15},
   issn = {1367-2630},
   issue = {10},
   journal = {New Journal of Physics},
   month = {10},
   pages = {103018},
   title = {Dissipative quantum Hopfield network: a numerical analysis},
   volume = {26},
   url = {https://iopscience.iop.org/article/10.1088/1367-2630/ad5e15},
   year = {2024}
}

@article{Lipkin1965a,
  author  = {Lipkin, H. J. and Meshkov, N. and Glick, A. J.},
  title   = {Validity of many-body approximation methods for a solvable model: I. Exact solutions and perturbation theory},
  year    = {1965},
  journal = {Nuclear Physics},
  volume  = {62},
  pages   = {188}
}

@article{Lipkin1965b,
  author  = {Lipkin, H. J. and Meshkov, N. and Glick, A. J.},
  title   = {Validity of many-body approximation methods for a solvable model: II. Linearization procedures},
  year    = {1965},
  journal = {Nuclear Physics},
  volume  = {62},
  pages   = {199}
}

@article{Lipkin1965c,
  author  = {Lipkin, H. J. and Meshkov, N. and Glick, A. J.},
  title   = {Validity of many-body approximation methods for a solvable model: III. Diagram summations},
  year    = {1965},
  journal = {Nuclear Physics},
  volume  = {62},
  pages   = {211}
}

@book{RingSchuck1980,
  author    = {Ring, Peter and Schuck, Peter},
  title     = {The Nuclear Many-Body Problem},
  address   = {New York},
  publisher = {Springer},
  year      = {1980}
}

@article{KitagawaUeda1993,
  author  = {Kitagawa, Masahiro and Ueda, Masahito},
  title   = {Squeezed spin states},
  year    = {1993},
  journal = {Physical Review A},
  volume  = {47},
  pages   = {5138}
}

@article{DusuelVidal2004,
  author  = {Dusuel, S. and Vidal, J.},
  title   = {Finite-size scaling exponents of the Lipkin-Meshkov-Glick model},
  year    = {2004},
  journal = {Physical Review Letters},
  volume  = {93},
  pages   = {237204}
}

@article{MilburnEtAl1997,
  author  = {Milburn, G. J. and Corney, J. and Wright, E. M. and Walls, D. F.},
  title   = {Quantum dynamics of an atomic Bose–Einstein condensate in a double-well potential},
  year    = {1997},
  journal = {Physical Review A},
  volume  = {55},
  pages   = {4318--4324}
}

@article{Leggett2001,
  author  = {Leggett, A. J.},
  title   = {Bose–Einstein condensation in the alkali gases: Some fundamental concepts},
  year    = {2001},
  journal = {Reviews of Modern Physics},
  volume  = {73},
  pages   = {307--356}
}

@article{MicheliEtAl2003,
  author  = {Micheli, A. and Jaksch, D. and Cirac, J. I. and Zoller, P.},
  title   = {Many-particle entanglement in two-component Bose–Einstein condensates},
  year    = {2003},
  journal = {Physical Review A},
  volume  = {67},
  pages   = {013607}
}

@article{Castanos2006PRB,
  author  = {Casta{\~n}os, O. and L{\'o}pez-Pe{\~n}a, R. and Hirsch, J. and L{\'o}pez-Moreno, E.},
  title   = {Classical and quantum phase transitions in the Lipkin–Meshkov–Glick model},
  year    = {2006},
  journal = {Physical Review B},
  volume  = {74},
  pages   = {104118}
}

@article{NagyRomera2012,
  author  = {Nagy, A. and Romera, E.},
  title   = {Fisher information, Rényi entropy power and quantum phase transition in the Dicke model},
  year    = {2012},
  journal = {Physica A},
  volume  = {391},
  pages   = {3650}
}

@article{RomeraCalixtoNagy2012,
  author  = {Romera, E. and Calixto, M. and Nagy, A.},
  title   = {Entropic uncertainty and the quantum phase transition in the Dicke model},
  year    = {2012},
  journal = {Europhysics Letters},
  volume  = {97},
  pages   = {20011}
}

@article{CalixtoRomeraDelReal2012,
  author  = {Calixto, M. and Romera, R. and del Real, R.},
  title   = {Parity-symmetry-adapted coherent states and entanglement in quantum phase transitions of vibron models},
  year    = {2012},
  journal = {Journal of Physics A: Mathematical and Theoretical},
  volume  = {45},
  pages   = {365301}
}

@article{CastanosCalixtoPerezBernalRomera2015,
  author  = {Casta{\~n}os, O. and Calixto, M. and P{\'e}rez-Bernal, F. and Romera, E.},
  title   = {Identifying the order of a quantum phase transition by means of Wehrl entropy in phase space},
  year    = {2015},
  journal = {Physical Review E},
  volume  = {92},
  pages   = {052106}
}

@article{CalixtoCastanosRomera2017_JSTAT,
  author  = {Calixto, Manuel and Casta{\~n}os, Octavio and Romera, Elvira},
  title   = {Entanglement and quantum phase diagrams of symmetric multi-qubit systems},
  journal = {Journal of Statistical Mechanics: Theory and Experiment},
  year    = {2017},
  volume  = {2017},
  number  = {10},
  pages   = {103103},
  doi     = {10.1088/1742-5468/aa8703}
}

@article{JurcevicEtAl2014,
  author  = {Jurcevic, P. and Lanyon, B. P. and Hauke, P. and Hempel, C. and Zoller, P. and Blatt, R. and Roos, C. F.},
  title   = {Quasiparticle engineering and entanglement propagation in a quantum many-body system},
  year    = {2014},
  journal = {Nature},
  volume  = {511},
  pages   = {202}
}

@article{RichermeEtAl2014,
  author  = {Richerme, P. and Gong, Z.-X. and Lee, A. and Senko, C. and Smith, J. and Foss-Feig, M. and Michalakis, S. and Gorshkov, A. V. and Monroe, C.},
  title   = {Non-local propagation of correlations in quantum systems with long-range interactions},
  year    = {2014},
  journal = {Nature},
  volume  = {511},
  pages   = {198}
}
\end{document}